\documentclass[aps,pra,twocolumn,groupedaddress,superscriptaddress,nofootinbib,notitlepage,floatfix,showpacs,amssymb,amsmath,amsfonts]{revtex4-2}

\usepackage{times,bm,bbm,bbold,amssymb,amsmath,amsfonts,dsfont,graphics,graphicx,braket,color,xcolor,hyperref}

\usepackage{amsmath}
\usepackage{algorithm,algpseudocode}

\usepackage{booktabs}

\graphicspath{{images/}}
\hypersetup{colorlinks,linkcolor={blue},citecolor={blue},urlcolor= {blue}}
\DeclareFontFamily{OT1}{pzc}{}
\DeclareFontShape{OT1}{pzc}{m}{it}
{<-> s * [1.25] pzcmi7t}{}
\DeclareMathAlphabet{\mathpzc}{OT1}{pzc}
{m}{it}

\newcommand{\ignore}[1]{}
\usepackage[T1]{fontenc}
\usepackage{newtxtext,newtxmath}

\begin{document}


\title{Novel Market Temperature Definition Through Fluctuation Theorem: A Statistical Physics Framework for Financial Crisis Prediction}

\author{Masoome Ramezani}
\affiliation{Department of Industrial Management, Faculty of Management and Economics, Science and Research Branch, Islamic Azad University, Tehran, Iran}

\author{Fereydoun Rahnama Roodposhti}
\affiliation{Department of Business Management, Faculty of Management and Economics, Science and Research Branch, Islamic Azad University, Tehran, Iran}

\author{Ghanbar Abbaspour Esfeden}
\affiliation{Department of Industrial Management, Faculty of Management, South Tehran Branch, Islamic Azad University, Tehran, Iran}

\author{Mehdi Ramezani}
\affiliation{Department of Physics, Sharif University of Technology, Tehran, Iran}

\date{\today}

\begin{abstract}
	\noindent This paper introduces a novel approach to financial crisis prediction by establishing a thermodynamic-like framework derived from the fluctuation theorem of statistical physics. We define market temperature through the probability ratio of positive to negative returns and demonstrate its effectiveness in identifying market states and predicting potential crises. Our empirical analysis spans nine major global indices from 2005 to 2025, revealing statistically significant differences in temperature dynamics between crisis and non-crisis periods. Most notably, we discover a counterintuitive relationship between market temperature stability and crisis occurrence: crises tend to emerge more frequently during periods of apparent temperature stability rather than instability. This finding suggests that unusually stable periods in market temperature might signal the accumulation of systemic risks, similar to the calm before a storm in physical systems.
\end{abstract}

\date{\today}
\maketitle

\section{INTRODUCTION}

Econophysics, emerging as a distinct interdisciplinary field in the 1990s, represents the application of methodologies and theories from statistical physics to understand complex economic systems \cite{mantegna1999introduction}. This field has grown from the recognition that financial markets and economic systems exhibit many characteristics familiar to physicists: complex interactions, emergent phenomena, phase transitions, and scaling laws \cite{bouchaud2003theory}. Just as physics describes the behavior of particles in a system, econophysics aims to understand the collective behavior of economic agents in markets \cite{farmer2009virtues}.

The field has made significant contributions to our understanding of financial markets, particularly in areas where traditional economic theories have shown limitations. While conventional economic models often assume rational behavior and equilibrium conditions \cite{fama1970efficient}, econophysics approaches markets as complex adaptive systems, where global patterns emerge from local interactions \cite{sornette2017stock}. This perspective has proven particularly valuable in explaining market phenomena that deviate from classical economic assumptions, such as fat-tailed distributions in returns, volatility clustering, and extreme events \cite{mantegna1999introduction,cont2001empirical}.

One of the most powerful aspects of the econophysics approach is its ability to identify and apply universal principles across seemingly different systems. Just as temperature serves as a fundamental measure in physical systems, characterizing the aggregate behavior of countless particles, similar aggregate measures might illuminate the collective behavior of market participants \cite{amaral1999econophysics, sornette2017stock}.

The concept of market temperature emerges as a natural extension of the statistical physics paradigm in financial markets. While traditional financial metrics like volatility provide some insight into market behavior, we propose a more fundamental approach based on the fluctuation theorem. This theorem, a cornerstone of non-equilibrium statistical mechanics, describes the probability ratio of entropy-producing and entropy-consuming processes in terms of temperature and heat exchange \cite{evans2002fluctuation}.

In physical systems, the fluctuation theorem states that the probability of observing a process that absorbs heat divided by the probability of observing a process that releases the same amount of heat is related exponentially to the amount of heat transfer and the temperature difference between the systems \cite{jarzynski2004classical,ramezani2018fluctuation,ramezani2018quantum}. By adapting this principle to financial markets, we can define a market temperature that captures not just the magnitude of price movements, but the fundamental asymmetry between market gains and losses.

This temperature metric provides a novel way to quantify market states, where deviations from normal conditions might signal impending instability. Unlike traditional volatility measures, our temperature definition through the fluctuation theorem captures the essential thermodynamic-like behavior of the market system.

Financial crises represent significant disruptions to economic systems, often resulting in substantial wealth destruction, unemployment, and social hardship that can persist long after the initial shock. The 2008 global financial crisis alone led to estimated losses exceeding \$10 trillion, demonstrating the devastating impact these events can have on both financial markets and the broader economy. Despite their profound consequences, our ability to predict and prevent financial crises remains limited, making the development of reliable early warning indicators a crucial challenge in modern economics.

Traditional approaches to crisis prediction often rely on conventional economic indicators such as credit-to-GDP ratios, asset price bubbles, or current account deficits. However, these metrics frequently fail to capture the complex, non-linear dynamics that characterize pre-crisis periods. Moreover, they typically signal potential problems either too late for preventive action or generate false alarms that could themselves trigger market instability. This limitation underscores the need for novel, physics-inspired approaches that can detect subtle changes in market dynamics before they manifest as full-blown crises.

The ability to accurately predict financial crises would not only allow policymakers to implement preventive measures but would also enable market participants to better manage risk and adjust their strategies accordingly. Furthermore, as financial markets become increasingly interconnected and complex, the potential for systemic risks has grown, making early warning systems more crucial than ever.

This paper introduces a novel approach to financial crisis prediction by establishing a thermodynamic-like framework based on the fluctuation theorem from statistical physics. Our key contribution is twofold. First, we develop a rigorous definition of market temperature derived from the fluctuation theorem, which relates the probability ratio of heat absorption to heat release in financial markets. This definition moves beyond traditional volatility measures to capture fundamental asymmetries in market behavior that may signal approaching instability.

Second, we demonstrate the practical utility of this temperature metric by developing a novel stability measure that identifies distinct market phases. Our analysis across major global indices reveals an unexpected pattern: financial crises tend to occur more frequently during periods of apparent temperature stability rather than instability. This counterintuitive finding suggests that market temperature stability might actually serve as an indicator of underlying systemic stress, much like the calm before a storm in physical systems. By examining temperature dynamics across different market states in indices spanning the Americas, Europe, and Asia-Pacific regions, we show that our approach captures subtle market behaviors that traditional volatility-based measures might miss. This framework not only advances our theoretical understanding of financial market dynamics but also offers practical tools for crisis prediction that could complement existing early warning systems.

The remainder of this paper is organized as follows. Section II establishes the theoretical framework, developing our market temperature definition from the fluctuation theorem and connecting it to established market patterns. Section III presents our empirical analysis of nine major global indices from 2005 to 2025, revealing that financial crises occur more frequently during periods of temperature stability rather than instability. Section IV discusses the implications of this counterintuitive finding for market monitoring and risk assessment. Finally, Section V concludes with a summary of our contributions and suggests directions for future research.


\section{THEORETICAL FRAMEWORK}
The application of physical theories to financial markets requires a careful foundation that bridges the gap between these seemingly disparate domains. To build this bridge systematically, we first examine the empirical evidence for predictable patterns in financial markets, documenting statistical regularities that suggest underlying universal principles. These patterns, ranging from power-law distributions in returns to long-range correlations in volatility, provide the empirical basis for applying physics-inspired approaches. We then explore how concepts from statistical physics have been adapted to understand these market phenomena, with particular attention to previous attempts at defining market temperature and their limitations. Finally, we present our novel approach based on the fluctuation theorem, showing how it provides a more fundamental and theoretically grounded definition of market temperature that captures essential features of market dynamics, particularly during periods of stress.

\subsection{Predictable Patterns in Financial Markets}
Financial markets, despite their apparent randomness, exhibit numerous robust statistical patterns that persist across different markets, time periods, and geographical regions. These patterns suggest underlying universal principles governing market behavior and provide the empirical foundation for applying physics-based approaches to financial systems.

The most fundamental and well-documented pattern appears in the distribution of returns. While early financial theories assumed Gaussian distributions, empirical evidence consistently shows that returns follow a more complex distribution characterized by heavy tails. The tails of the cumulative distribution follow a power law: $\Pr(|R| > x) \sim x^{-\alpha}$ with $\alpha \approx 3 $, a phenomenon known as the inverse cubic law \cite{gopikrishnan1998inverse}. This pattern has been observed across diverse markets, from mature indices like the S\&P 500 to emerging market exchanges, and persists across different time scales from minutes to months \cite{gopikrishnan1999scaling,lux1996stable,pan2008inverse}. The center of the distribution shows approximately exponential behavior, with a transition to power-law tails occurring at roughly three standard deviations from the mean.

Temporal correlations in market activity present another set of robust patterns \cite{mandelbrot1997variation, granger1995some,ding1996modeling, cont2007volatility}. While returns themselves show minimal serial correlation, consistent with market efficiency, their magnitude exhibits strong temporal dependencies. This phenomenon, known as volatility clustering, manifests as a power-law decay \cite{dubrulle1997scale,guillaume1997bird} in the autocorrelation function of absolute returns: $C(\tau) \sim \tau^{-\gamma}$ with $\gamma \le 0.5$ \cite{dubrulle1997scale,baillie1996long}. In practical terms, this means that large price changes tend to be followed by other large changes, and periods of relative calm tend to persist, creating distinct market regimes that can extend over several months.

Trading volume patterns provide additional evidence of market regularities. The distribution of trading volumes follows a power law with an exponent near 1.5 \cite{gopikrishnan2000statistical,gabaix2003theory}, indicating scale-invariant behavior in market activity. More sophisticated analysis reveals intricate relationships between volume and price movements: the average volume for a given price change $\Delta P$ scales as $\langle V \rangle \sim |\Delta P|^{\beta}$ with $\beta \approx 2$ \cite{gabaix2003theory}. These relationships hold remarkably constant across different market conditions, suggesting fundamental constraints on market dynamics.

The microstructure of markets, revealed through order book dynamics, displays its own set of regularities. Limit order sizes follow a power-law distribution with an exponent approximately -2.5 \cite{maslov2001price}, while the spread between best bid and ask prices exhibits complex scaling behavior reminiscent of critical phenomena in physical systems. The flow of orders shows long-range correlations similar to those observed in turbulent flows, suggesting possible analogies with physical systems far from equilibrium.

Studies of both U.S. tax-paying firms \cite{axtell2001zipf} and Japanese companies \cite{okuyama1999zipf} have demonstrated that the distribution of firm sizes and annual income follows Zipf's law, where the probability of a firm being larger than size s is inversely proportional to s.

Analysis of stock market price changes using random matrix theory revealed universal patterns in financial markets, as demonstrated by the cross-correlation matrix of 1000 US stocks exhibiting properties consistent with theoretical predictions of random matrices \cite{plerou1999universal}.

Intraday patterns add another dimension to market regularities. Trading activity typically follows a U-shaped pattern during the trading day, with higher volumes at the opening and closing of markets. The distribution of times between trades follows a stretched exponential function, and trading activity patterns show self-similar behavior across different time scales, suggesting fractal market structure.

These diverse but interconnected patterns paint a picture of financial markets as complex systems operating near a critical state. The universality of these patterns across different markets and time periods suggests the existence of fundamental principles governing market behavior, much like physical laws govern the behavior of matter. This empirical foundation motivates the application of statistical physics methods to understand and predict market dynamics, particularly during periods of stress or transition.

\subsection{Statistical Physics in Financial Markets}
The remarkable parallels between financial market behavior and physical systems extend beyond surface-level analogies, providing a rigorous framework for understanding market dynamics. Statistical physics, with its focus on emergent behavior arising from interactions among many individual components, offers powerful tools for analyzing financial markets as complex adaptive systems.

The foundation of this approach lies in recognizing that financial markets, like physical systems, exhibit collective behavior that cannot be reduced to the simple aggregation of individual actions. Just as the macroscopic properties of gases emerge from the collective motion of molecules, market-wide phenomena emerge from the interactions of numerous traders, each making decisions based on limited information and their own objectives.

One of the most successful applications of statistical physics to financial markets appears in the analysis of phase transitions \cite{kasprzak2010higher}. Market crashes, for instance, share striking similarities with critical phenomena in physical systems. The increasing correlation length between different stocks leading up to a market crash mirrors the behavior of physical systems approaching a critical point, where local fluctuations become increasingly coordinated. This phenomenon manifests in the power-law scaling of correlation functions and susceptibility measures, suggesting that markets operate near a critical state.

The concept of universality, central to statistical physics, finds particular resonance in financial markets \cite{mantegna1999introduction, bouchaud2003theory}. Just as different physical systems can exhibit identical critical exponents despite having different microscopic dynamics, financial markets across different regions and time periods show remarkably consistent scaling laws. The inverse cubic law of returns and the power-law decay of volatility autocorrelations exemplify this universality, suggesting fundamental organizing principles independent of specific market mechanisms.

Entropy concepts from statistical physics have proven especially valuable in understanding market efficiency and information flow \cite{zhou2013applications}. The efficient market hypothesis can be reframed in terms of maximum entropy production, where market prices evolve to maximize information entropy under constraints imposed by available information and trading mechanisms. This perspective helps explain why markets tend to exhibit both periods of apparent efficiency and systematic deviations from random walk behavior.

The analogy between energy in physical systems and money in financial markets provides another fertile ground for applying statistical physics concepts \cite{mccauley2003thermodynamic}. Trading activities can be viewed as energy exchanges, with price formation emerging from the collective interaction of these exchanges, similar to how temperature emerges from energy exchanges between particles. However, this analogy must be treated carefully, as financial markets exhibit several unique features not present in physical systems, particularly the role of expectations and strategic behavior.

Renormalization group methods, originally developed to understand critical phenomena in physics, have found applications in analyzing the hierarchical structure of financial markets \cite{wu2012critical}. These techniques help explain how interactions across different time scales give rise to the observed scaling properties of financial time series, from intraday patterns to long-term market trends. The success of these methods suggests that market dynamics possess a degree of scale invariance, similar to critical systems in physics.

Most significantly for our current work, the fluctuation-dissipation relations from statistical physics provide a theoretical framework for understanding market responses to external perturbations \cite{yura2014financial}. These relations suggest that the same underlying mechanisms govern both the response to external forces and the natural fluctuations in the system, a principle that proves crucial in developing our market temperature metric.

The application of statistical physics to financial markets extends beyond mere analogy, offering quantitative tools for analyzing market stability and predicting critical transitions. However, traditional approaches have often focused on equilibrium concepts, despite clear evidence that financial markets operate far from equilibrium. Our work addresses this limitation by applying non-equilibrium statistical physics, particularly the fluctuation theorem, to develop a more comprehensive understanding of market dynamics.

This statistical physics framework suggests that market crashes and other critical events might be understood not as external shocks but as intrinsic features of the market's dynamics, emerging from the collective behavior of interacting agents. This perspective motivates our development of a market temperature metric based on fundamental physical principles rather than ad hoc financial indicators.

\subsection{The Fluctuation Theorems and Market Temperature}
The story of fluctuation theorems represents one of the most profound intellectual achievements in statistical physics, providing a mathematical bridge between microscopic reversibility and macroscopic irreversibility - a puzzle that had challenged physicists since Boltzmann's time. To appreciate its significance, let's trace its development through a series of groundbreaking discoveries.

The modern journey began in 1993 when Evans, Cohen, and Morriss made a remarkable observation while studying sheared fluids \cite{evans1993probability}. They discovered that even in systems driven far from equilibrium, there existed a precise mathematical relationship between entropy-producing and entropy-consuming trajectories:
\begin{equation}
	\frac{P(+\sigma)}{P(-\sigma)} = \exp(\tau \sigma),
\end{equation}
where $P(\pm\sigma)$ represents the probability of observing an entropy production rate $\pm\sigma$ over time $\tau$. This elegant relation showed that while the Second Law of Thermodynamics holds on average, violations could occur at small scales and short times with precisely quantifiable probabilities.
	
Building on this foundation, Jarzynski made a revolutionary contribution in 1997 with his equality \cite{jarzynski1997nonequilibrium}:
\begin{equation}
	\langle \exp(-\beta W) \rangle = \exp(-\beta \Delta F),
\end{equation}
Here, $W$ represents the work done on the system, $\beta$ is the inverse temperature $(1/kT)$, and $\Delta F$ is the free energy difference between initial and final equilibrium states. The angular brackets $\langle ... \rangle$ denote an average over all possible trajectories. Think of this as a bridge between two worlds: the messy reality of non-equilibrium processes and the clean, predictable world of equilibrium thermodynamics. Just as a suspension bridge connects two landmasses while floating free of both, the Jarzynski equality connects equilibrium properties to non-equilibrium measurements without requiring the system to stay close to equilibrium.

Crooks further expanded this framework in 1999 with his fluctuation theorem \cite{crooks1999entropy}:
\begin{equation}
	\frac{P_F(+W)}{P_R(-W)} = \exp(\beta (W-\Delta F)).
\end{equation}
This relation connects the probability $P_F(+W)$ of observing work $+W$ in a forward process to the probability $P_F(-W)$ of observing work $-W$ in the reverse process. 

\begin{figure*}[!htbp]
	\includegraphics[width=0.3\textwidth]{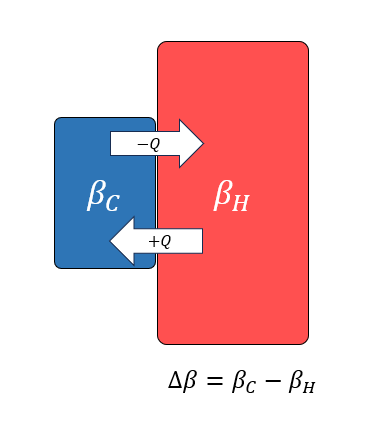}
	\caption{Schematic representation of heat exchange between two systems at different temperatures}
	\label{fig:fluctuation}
\end{figure*}

The framework reached a new level of universality with Jarzynski and Wójcik's 2004 quantum fluctuation theorem for heat exchange \cite{jarzynski2004classical,ramezani2018fluctuation,ramezani2018quantum}:
\begin{equation}
	\frac{P(+Q;\tau)}{P(-Q;\tau)} = \exp(Q \Delta\beta).
\end{equation}
This relation, which forms the theoretical foundation of our work, describes heat exchange between systems with inverse temperature difference $\Delta\beta$. To understand this intuitively, consider two systems at different temperatures, as illustrated in Fig. \ref{fig:fluctuation}. The cold system, characterized by inverse temperature $\beta_C$, exchanges heat with a hot system at inverse temperature $\beta_H$. The difference $\Delta \beta = \beta_C - \beta_H $ acts like a thermodynamic voltage, driving heat flow between the systems. When $\Delta \beta$ is positive, processes transferring heat from hot to cold become exponentially more likely than their reverse processes, though importantly, the theorem still allows for such reverse processes with precisely quantified probabilities.

We propose that this framework translates remarkably well to financial markets. Instead of physical systems exchanging heat, we interpret $Q$ as returns, representing the transfer of value between market participants. The time interval $\tau$ becomes our observation window for calculating the probability of positive and negative returns. This analogy runs deeper than mere mathematical convenience - in both cases, the exponential relationship emerges from the collective behavior of many interacting components operating under constraints.

To operationalize this insight, we plot $\log(P(+Q)/P(-Q))$ versus $Q$ using a time window $\tau$. The slope of this relationship $\Delta \beta$, provides our definition of market temperature. This approach mirrors how physicists measure temperature through the relationship between energy flows and probability ratios, rather than through direct observation of molecular motion. Just as the temperature difference  $\Delta \beta$ in physical systems drives the direction of heat flow, in markets it represents an imbalance in market conditions that drives directional movement in prices.

\subsection{Crisis in Financial Markets}
The detection and prediction of financial crises represent a central challenge in modern economics, requiring both precise definition and robust quantification methods. While traditional approaches often rely on threshold-based measures of market volatility or drawdown, we propose a more systematic method based on the CMAX function, which provides a normalized measure of market decline relative to previous peaks \cite{patel1998crises}.

Let $X = \{x_t, t = 1,2, ..., N\}$ represent the time series of a market index. We define the CMAX function as ${C_{\text{max}} = \{c_{\text{max}}(t), t = T,T+1, ..., N\}}$, where:
\begin{equation}
	c_{\text{max}}(t) = \frac{x_{t}}{\max[\{x_{t'}|t'=t-T,t-T+1,...,t\}]},
\end{equation}
here, $T$ represents the lookback window, typically set to either 360 or 720 trading days, providing sufficient historical context while remaining responsive to current market conditions. The CMAX function effectively normalizes current market levels against their recent maximum, producing a metric bounded between 0 and 1, where values closer to 1 indicate price levels near historical highs and lower values suggest significant market declines.

Building on this foundation, we formalize our crisis identification framework through a binary classification function:
\begin{equation}
	\text{crisis}(t) = \begin{cases} 
		1, & c_{\text{max}}(t) < \mu - n\sigma \\
		0, & \text{otherwise}
	\end{cases}
\end{equation}
where $\mu$ and $\sigma$ represent the mean and standard deviation of the CMAX time series respectively. In our analysis, we set $n=1$, identifying crisis periods as those where the CMAX value falls more than one standard deviation below its historical mean. This calibration provides a balanced approach to crisis detection, capturing significant market stress events while minimizing false positives.

The crisis indicator $\text{crisis}(t) = 1$ identifies periods where market behavior deviates significantly from normal conditions, providing a quantitative basis for studying the relationship between market temperature dynamics and the onset of financial crises. This relationship forms the foundation for our predictive framework, which we develop in subsequent sections.

\section{RESULTS AND ANALYSIS}
Our empirical analysis reveals several striking patterns in market temperature dynamics across major global indices during the period from January 2005 to January 2025. By applying the fluctuation theorem framework developed in Section II to real market data, we uncover a counterintuitive relationship between market stability and crisis occurrence. The analysis encompasses nine major indices across the Americas, Europe, and Asia-Pacific regions, providing a comprehensive test of our theoretical predictions. Using a rolling window approach to measure market temperature stability and a CMAX-based crisis identification method, we find statistically significant differences in temperature dynamics between crisis and non-crisis periods for most indices. Most notably, our results demonstrate that periods of apparent temperature stability often precede market crises, suggesting that market calm may paradoxically signal underlying systemic stress. The following subsections present a detailed examination of these findings, beginning with the fundamental temperature dynamics across different market states, proceeding through our stability threshold methodology, and culminating in a comprehensive analysis of crisis occurrence patterns and regional variations.

\subsection{Market Temperature Dynamics}

\begin{table*}[htbp]
	\centering
	\caption{Market Temperature Statistics and Statistical Tests for Crisis and Non-Crisis Periods}
	\footnotesize
	\begin{tabular}{lcccccc}
		\toprule
		& \multicolumn{2}{c}{$\Delta\beta$ in Crisis Period} & \multicolumn{2}{c}{$\Delta\beta$ in Non-Crisis Period} & \multicolumn{2}{c}{Mann-Whitney U Test} \\
		\cmidrule(lr){2-3} \cmidrule(lr){4-5} \cmidrule(lr){6-7}
		Market & Mean & Std & Mean & Std & U statistic & p-value \\
		\midrule
		\multicolumn{7}{l}{\textit{Americas Indices}} \\
		S\&P 500   & -10.8019 & 3.2003 & -13.9565 & 10.1179 & 1280501.0000 & 0.0000 \\
		Dow Jones  & -8.2114  & 3.7301 & -11.5100 & 11.9263 & 1161522.5000 & 0.0001 \\
		Nasdaq 100 & -12.0926 & 4.1660 & -10.3564 & 13.6559	& 1184808.5000 & 0.8632 \\
		\midrule
		\multicolumn{7}{l}{\textit{European Indices}} \\
		DAX           & -12.2606 & 7.7211  & -7.6685 & 14.3394 & 1043035.5000 & 0.0000 \\
		FTSE 100      & -8.7725  & 14.1149 & -9.8064 & 9.7878  & 1421032.5000 & 0.0000 \\
		Euro Stoxx 50 & -8.2622	 & 10.2238 & -5.5545 & 9.5331  & 497669.5000  &	0.0020\\
		\midrule
		\multicolumn{7}{l}{\textit{Asia \& Pacific Indices}} \\
		Hang Seng    & -17.2386 & 4.4255 & -7.5672	& 10.0668 & 1204167.5000 & 0.1542 \\
		S\&P/ASX 200 & 3.9238	& 6.2936 & -13.6102	& 16.3393 & 1889566.0000 & 0.0000 \\
		Nifty 50     & -3.8585	& 6.7860 & -3.4254	& 9.8046  & 897135.0000	 & 0.0050 \\
		\bottomrule
	\end{tabular}
	\label{tab:market_temperature}
\end{table*}

The application of our fluctuation theorem-based temperature metric $(\Delta \beta)$ reveals distinct patterns between crisis and non-crisis periods across global financial markets. Table \ref{tab:market_temperature} presents a comprehensive statistical comparison of market temperature dynamics across nine major global indices, including mean values, standard deviations, and Mann-Whitney U test results for both crisis and non-crisis periods.

Our analysis demonstrates that market temperature exhibits significantly different behavior during periods of market stress compared to normal trading conditions. As shown in Table \ref{tab:market_temperature}, the S\&P 500 exhibits a mean temperature differential of -10.8019 $(\sigma = 3.2003)$ during crisis periods, contrasting with -13.9565 $(\sigma = 10.1179)$ during non-crisis periods. This difference is highly statistically significant ${(U = 1280501, p < 0.0001)}$, indicating fundamentally different market dynamics between these states.

The pattern of distinct temperature regimes extends across most major global indices, as evidenced by the data in Table \ref{tab:market_temperature}. The Dow Jones Industrial Average shows similar behavior, with crisis periods characterized by ${\Delta \beta = -8.2114}$ ${(\sigma = 3.7301)}$ compared to -11.5100 $(\sigma = 11.9263 )$ during non-crisis periods ${(U = 1161522.5, p < 0.0001)}$ . European markets demonstrate even more pronounced differences, with the DAX exhibiting a crisis period temperature of -12.2606 $(\sigma = 7.7211 )$ versus -7.6685 $(\sigma = 14.3394 )$ in non-crisis periods ${(U = 1043035.5, p < 0.0001)}$.

Particularly noteworthy in Table \ref{tab:market_temperature} is the consistency in statistical significance across different market environments. Seven of the nine analyzed indices show highly significant differences $(p < 0.05)$ in temperature dynamics between crisis and non-crisis periods, suggesting that our temperature metric captures a fundamental aspect of market behavior that transcends regional and structural differences in trading systems. The magnitude of these differences, coupled with their statistical significance, provides strong empirical support for the theoretical framework developed in Section II.

The Asia-Pacific markets present interesting variations in this pattern, as clearly shown in Table \ref{tab:market_temperature}. The S\&P/ASX 200 stands out with a positive mean temperature during crisis periods ${(\Delta \beta = 3.9238, \sigma = 6.2936)}$, contrasting sharply with its negative non-crisis temperature ${(\Delta \beta = -13.6102, \sigma = 16.3393)}$. This unique behavior suggests that while the temperature metric effectively distinguishes between market states across all regions, the specific characteristics of these states may vary with market structure and regional factors.

The robustness of these findings is further supported by the Mann-Whitney U test results presented in the rightmost columns of Table \ref{tab:market_temperature}, which account for potential non-normality in the temperature distributions. The consistently low p-values across most indices indicate that the observed differences are not artifacts of outliers or specific distributional assumptions but rather reflect genuine distinctions in market behavior during different states. These results establish a strong empirical foundation for our subsequent analysis of market stability and crisis prediction.

\subsection{Temperature Stability and Crisis Analysis}

\begin{table*}[htbp]
	\centering
	\caption{Market Temperature Stability Analysis and Crisis Occurrence Patterns}
	\begin{tabular}{lccccc}
		\toprule
		& Number of & Number of & Stability & \multicolumn{2}{c}{Crisis Occurrence (\%)} \\
		\cmidrule(lr){5-6}
		Market & Crisis Days & Stable Days & Threshold & Stable Periods & Unstable Periods \\
		\midrule
		\multicolumn{6}{l}{\textit{Americas Indices}} \\
		S\&P 500 & 606 & 1058 & 0.2195 & 18.24 & 12.82 \\
		Dow Jones & 568 & 1058 & 0.2203 & 20.42 & 10.93 \\
		\midrule
		\multicolumn{6}{l}{\textit{European Indices}} \\
		DAX & 666 & 1058 & 0.2633 & 20.70 & 13.88 \\
		FTSE 100 & 665 & 1058 & 0.1994 & 16.35 & 15.27 \\
		Euro Stoxx 50 & 490 & 668 & 0.2683 & 16.62 & 18.47 \\
		\midrule
		\multicolumn{6}{l}{\textit{Asia \& Pacific Indices}} \\
		S\&P/ASX 200 & 652 & 1058 & 0.3018 & 26.75 & 11.46 \\
		Nifty 50 & 522 & 1048 & 0.4132 & 10.59 & 12.87 \\
		\bottomrule
	\end{tabular}
	\label{tab:stability_analysis}
\end{table*}

Building upon the distinct temperature behaviors identified during crisis and non-crisis periods in Section III.A, we investigate how temperature stability relates to crisis occurrence. Our methodology employs rolling window statistics to quantify stability and examine its relationship with market crises.

For each trading day $t$, we compute both the rolling standard deviation $\sigma_{r}(t)$ and rolling mean $\mu_{r}(t)$  over a 50-day window, capturing the dynamic nature of market temperature variations. The rolling standard deviation, which serves as our primary stability indicator, is calculated as:
\begin{equation}
	\sigma_{r}^{2}(t) = \frac{1}{w} \sum_{i=t-w+1}^{t}[\Delta\beta(i) - \mu_{r}(t)]^{2},
\end{equation}
where $w = 50$ is the window size and $\mu_{r}(t)$ is the rolling mean over the same window.

We define the stability threshold $\Theta$ adaptively for each market as the 25th percentile of the rolling standard deviation distribution:
\begin{equation}
	\Theta = P_{25}(\sigma_{r}).
\end{equation}

This adaptive threshold approach reveals significant variations across markets, as shown in Table \ref{tab:stability_analysis}. The FTSE 100 exhibits a stability threshold of 0.1994, while the Nifty 50 shows a markedly higher threshold of 0.4132, reflecting fundamental differences in market dynamics.

Our analysis of crisis occurrence during stable periods $\sigma_{r}(t) \le \Theta$ versus unstable periods reveals a striking pattern. As Table \ref{tab:stability_analysis} demonstrates, most markets experience a higher percentage of crises during periods of apparent stability. The S\&P 500 experiences 18.24\% of its crisis periods during stable periods, compared to 12.82\% during unstable periods. This pattern is even more pronounced in the S\&P/ASX 200, where 26.75\% of crises occur during stable periods versus 11.46\% during unstable periods.

This tendency persists across different markets, with the Dow Jones showing 20.42\% of crises during stable periods compared to 10.93\% during unstable periods, and the DAX exhibiting 20.70\% versus 13.88\%. The Euro Stoxx 50 presents a notable exception, with crisis occurrences slightly higher during unstable periods (18.47\%) than stable periods (16.62\%), suggesting potential regional variations.

These findings indicate that periods of stable market temperature, rather than signaling market health, may indicate accumulating systemic stress. This observation has significant implications for market monitoring and risk assessment frameworks.

To provide a comprehensive view of these dynamics, detailed visualizations for each studied index are presented in Appendix A. These figures combine the market index values, temperature measurements, stability indicators, CMAX values, and crisis periods in a unified format, allowing readers to observe the interplay between temperature stability and crisis occurrence across different markets and time periods.

\section{DISCUSSION}
Our findings reveal a compelling relationship between market temperature dynamics and financial crises, with several noteworthy implications for both theoretical understanding and practical applications in market monitoring. The results presented in Section III demonstrate that the fluctuation theorem framework provides valuable insights into market behavior, particularly in identifying precursors to crisis events.

The most striking finding is the systematic difference in market temperature $(\Delta\beta)$ between crisis and non-crisis periods across major global indices. The statistical significance of these differences, supported by Mann-Whitney U tests, suggests that our temperature metric captures fundamental changes in market dynamics during periods of stress. However, two notable exceptions emerge from this pattern: the Nasdaq 100 and Hang Seng indices. The absence of significant temperature differences in these markets may reflect their unique characteristics. The Nasdaq 100, dominated by technology companies, often exhibits different behavior from traditional indices due to its sector concentration and growth orientation. Similarly, the Hang Seng's distinctive behavior might be attributed to the unique regulatory environment and market structure of the Hong Kong exchange.

Perhaps the most counterintuitive result is the higher occurrence of crises during periods of temperature stability. This "calm before the storm" phenomenon suggests that market stability, rather than indicating robustness, may signal the buildup of hidden systemic risks. This finding aligns with Minsky's Financial Instability Hypothesis, which posits that periods of stability encourage increased risk-taking and leverage, ultimately leading to instability. Our temperature metric provides a quantitative framework for identifying such periods of deceptive calm.

The variation in stability thresholds across markets is particularly informative. The significantly higher threshold observed in the Nifty 50 (0.4132) compared to the FTSE 100 (0.1994) suggests that emerging markets may require different stability criteria than developed markets. This observation has important implications for risk management practices, indicating that universal thresholds may be inappropriate for global market monitoring.

The Euro Stoxx 50's unique behavior, showing higher crisis occurrence during unstable periods, warrants special attention. This exception to the general pattern might reflect the complexities of the Eurozone market, where multiple national economies influence index behavior. The divergence suggests that regional economic integration may affect how market temperature relates to crisis development.

Our findings have significant implications for market surveillance and regulation. Traditional volatility-based monitoring systems might miss crucial signals by focusing on market turbulence rather than unusual stability. The temperature metric's ability to identify potentially dangerous stable periods offers a new tool for regulatory oversight. However, implementation would require careful consideration of market-specific characteristics, as evidenced by the varying stability thresholds across indices.

These results also contribute to the broader discussion of market efficiency and stability. The consistent relationship between temperature dynamics and crisis occurrence across most markets suggests that our framework captures universal aspects of market behavior, transcending regional and structural differences. This universality strengthens the case for applying statistical physics concepts to financial markets while acknowledging the need for market-specific calibration.

\section{CONCLUSION}
This paper introduces a novel approach to financial crisis prediction by establishing a thermodynamic-like framework based on the fluctuation theorem from statistical physics. Our investigation reveals that market temperature, derived from the probability ratio of positive to negative returns, provides a powerful tool for understanding market dynamics and anticipating potential crises. The empirical analysis across nine major global indices from 2005 to 2025 demonstrates that this framework captures subtle but significant patterns in market behavior that traditional measures might overlook.

The key contribution of our work lies in the discovery that periods of apparent market temperature stability often precede crisis events. This counterintuitive finding challenges conventional wisdom about market stability and suggests that unusually calm periods in market temperature might signal the accumulation of systemic risks. The statistical significance of temperature differences between crisis and non-crisis periods across most studied indices validates the fundamental premise of our approach, while the variations in stability thresholds across different markets highlight the importance of market-specific calibration.

Our framework's ability to identify potential crisis periods through temperature stability rather than volatility represents a significant advancement in market monitoring capabilities. This approach offers regulatory bodies and market participants a new perspective on risk assessment, complementing existing tools with a physics-inspired metric that captures fundamental market dynamics.

Looking forward, several promising directions for future research emerge from our findings. First, investigating the microscopic mechanisms that link temperature stability to crisis formation could deepen our understanding of market dynamics. Second, extending this framework to individual stocks and other financial instruments might reveal sector-specific patterns in temperature behavior. Finally, developing real-time monitoring systems based on our temperature metric could enhance early warning capabilities for market participants and regulators.

As financial markets continue to evolve and become increasingly interconnected, the need for sophisticated tools to monitor and predict potential crises becomes more crucial. Our thermodynamic framework provides both theoretical insights into market behavior and practical tools for crisis prediction, contributing to the ongoing effort to better understand and manage financial market risks.

\section*{CODE AVAILABILITY}
The complete implementation of our analysis framework, including the market temperature calculation and crisis prediction methodology, is available at \href{https://github.com/MehdiRamezani200/temperature-in-financial-markets}{https://github.com/MehdiRamezani200/temperature-in-financial-markets}.


\bibliographystyle{unsrt}
\bibliography{bibliography}


\onecolumngrid
\newpage

\appendix
\section*{Appendix A: Market Temperature Analysis Visualizations}
\addcontentsline{toc}{section}{Appendix A: Market Temperature Analysis Visualizations}

\subsection*{A1. S\&P 500}
\begin{figure}[h]
	\centering
	\includegraphics[width=\textwidth]{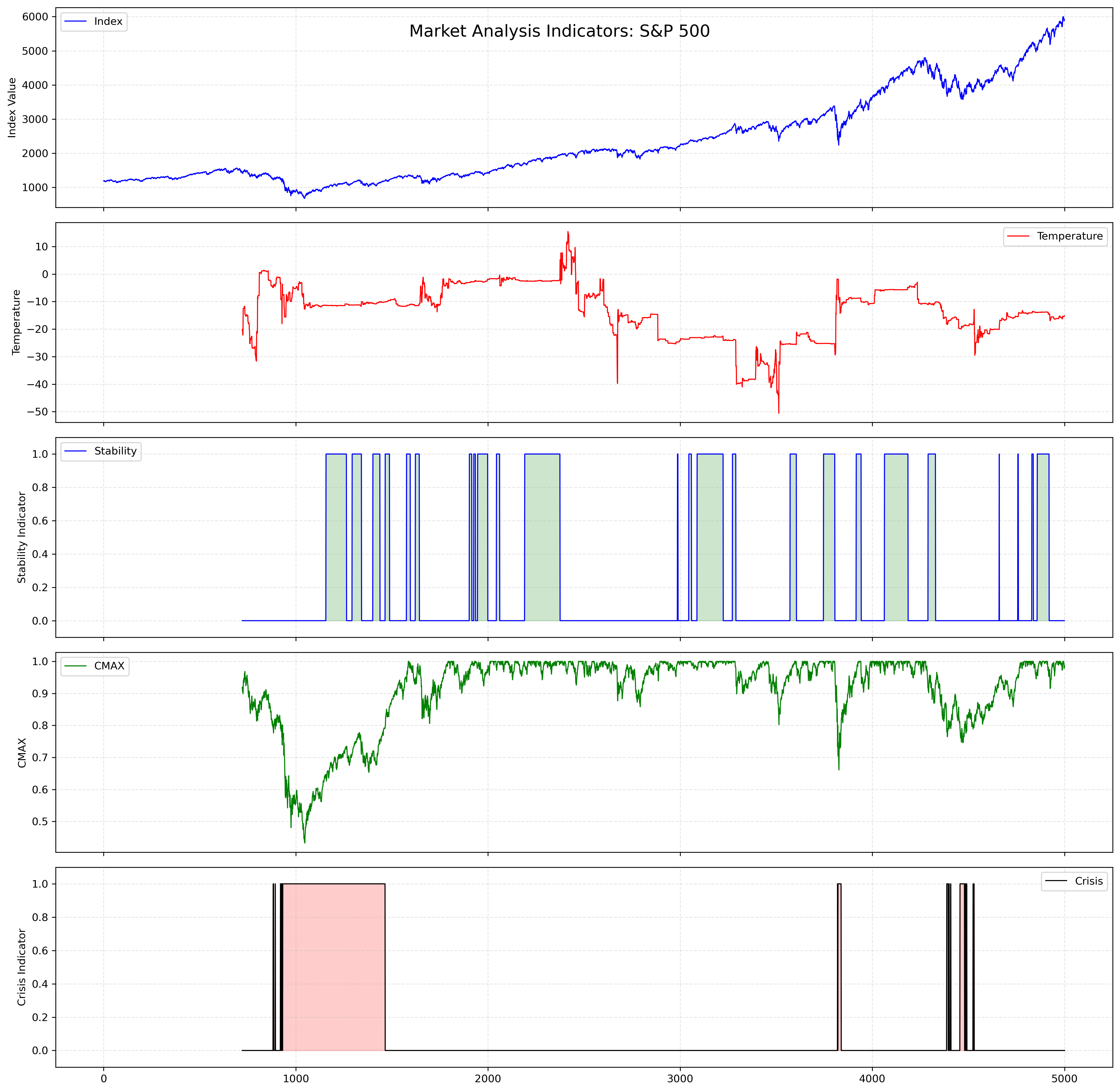}
	\caption{Market Analysis Indicators for S\&P 500. From top to bottom: (1) Index value over time, (2) Market temperature ($\Delta\beta$), (3) Stability indicator showing periods of temperature stability (blue shading), (4) CMAX values showing market drawdowns, and (5) Crisis indicator showing identified crisis periods (red shading).}
	\label{fig:SP500_analysis}
	\clearpage
\end{figure}

\newpage

\subsection*{A2. Dow Jones}
\begin{figure}[h]
	\centering
	\includegraphics[width=\textwidth]{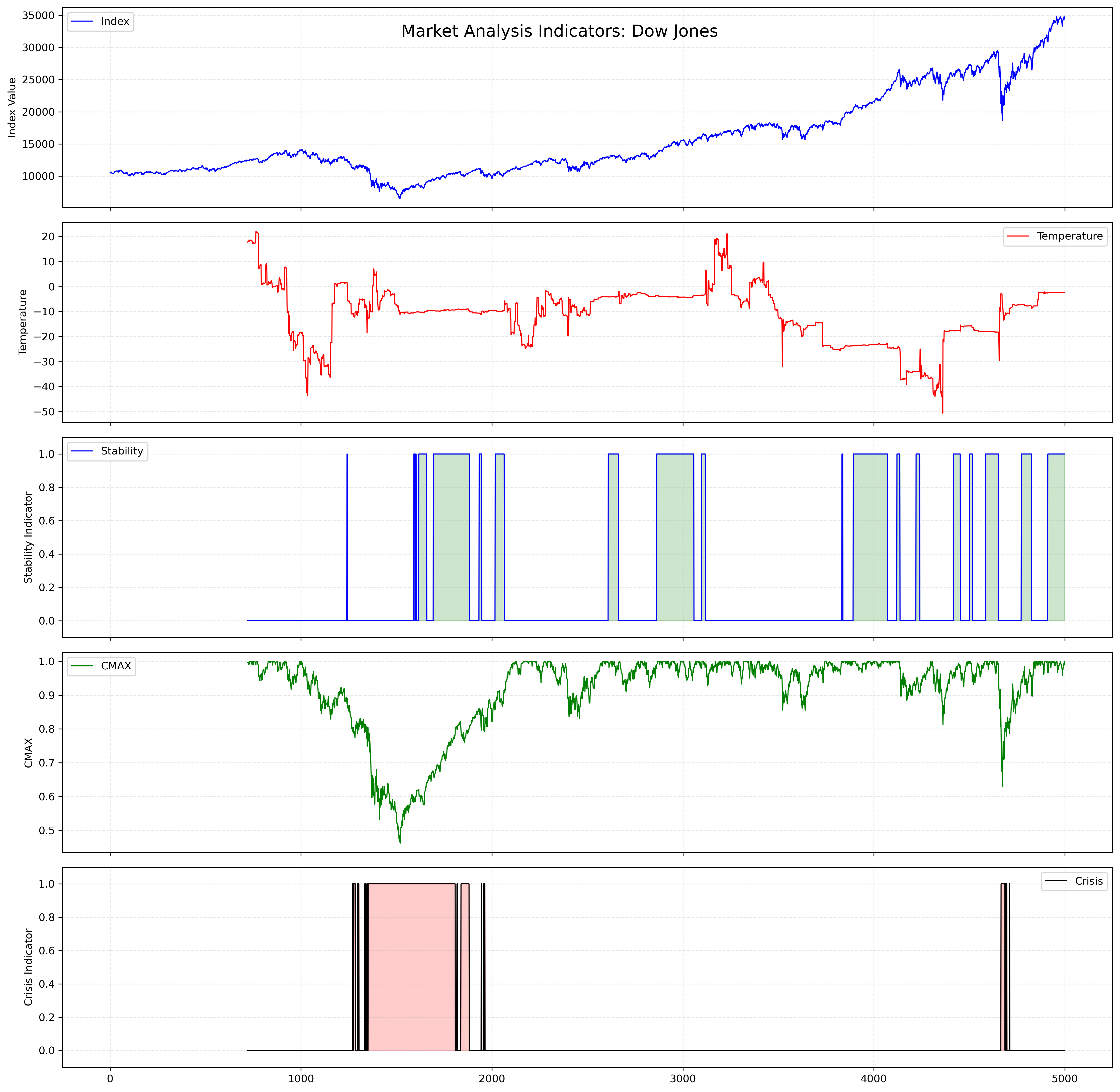}
	\caption{Market Analysis Indicators for S\&P 500. From top to bottom: (1) Index value over time, (2) Market temperature ($\Delta\beta$), (3) Stability indicator showing periods of temperature stability (blue shading), (4) CMAX values showing market drawdowns, and (5) Crisis indicator showing identified crisis periods (red shading).}
	\label{fig:DowJones_analysis}
	\clearpage
\end{figure}

\newpage

\subsection*{A3. Nasdaq 100}
\begin{figure}[h]
	\centering
	\includegraphics[width=\textwidth]{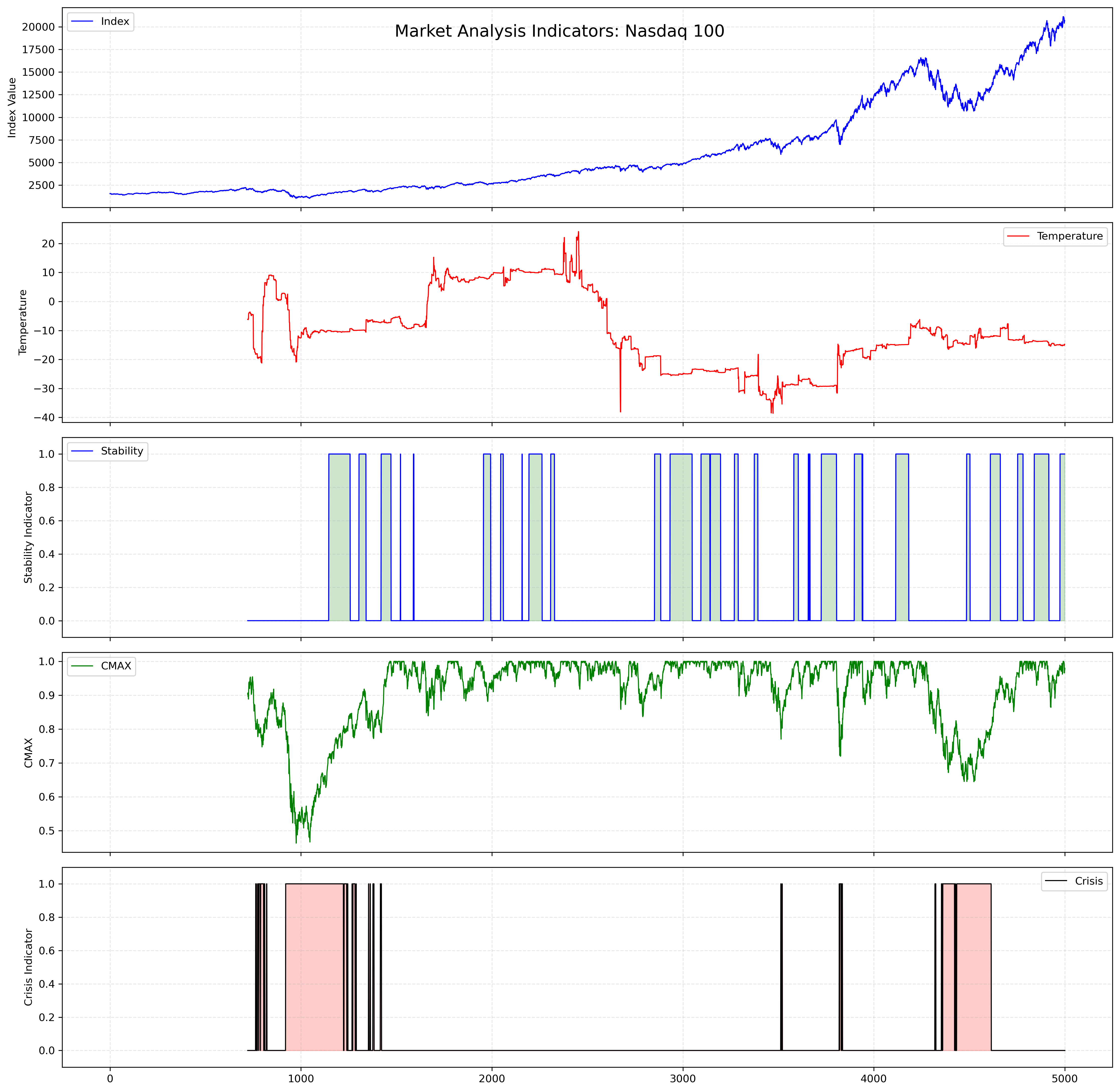}
	\caption{Market Analysis Indicators for S\&P 500. From top to bottom: (1) Index value over time, (2) Market temperature ($\Delta\beta$), (3) Stability indicator showing periods of temperature stability (blue shading), (4) CMAX values showing market drawdowns, and (5) Crisis indicator showing identified crisis periods (red shading).}
	\label{fig:Nasdaq100_analysis}
	\clearpage
\end{figure}

\newpage

\subsection*{A4. DAX}
\begin{figure}[h]
	\centering
	\includegraphics[width=\textwidth]{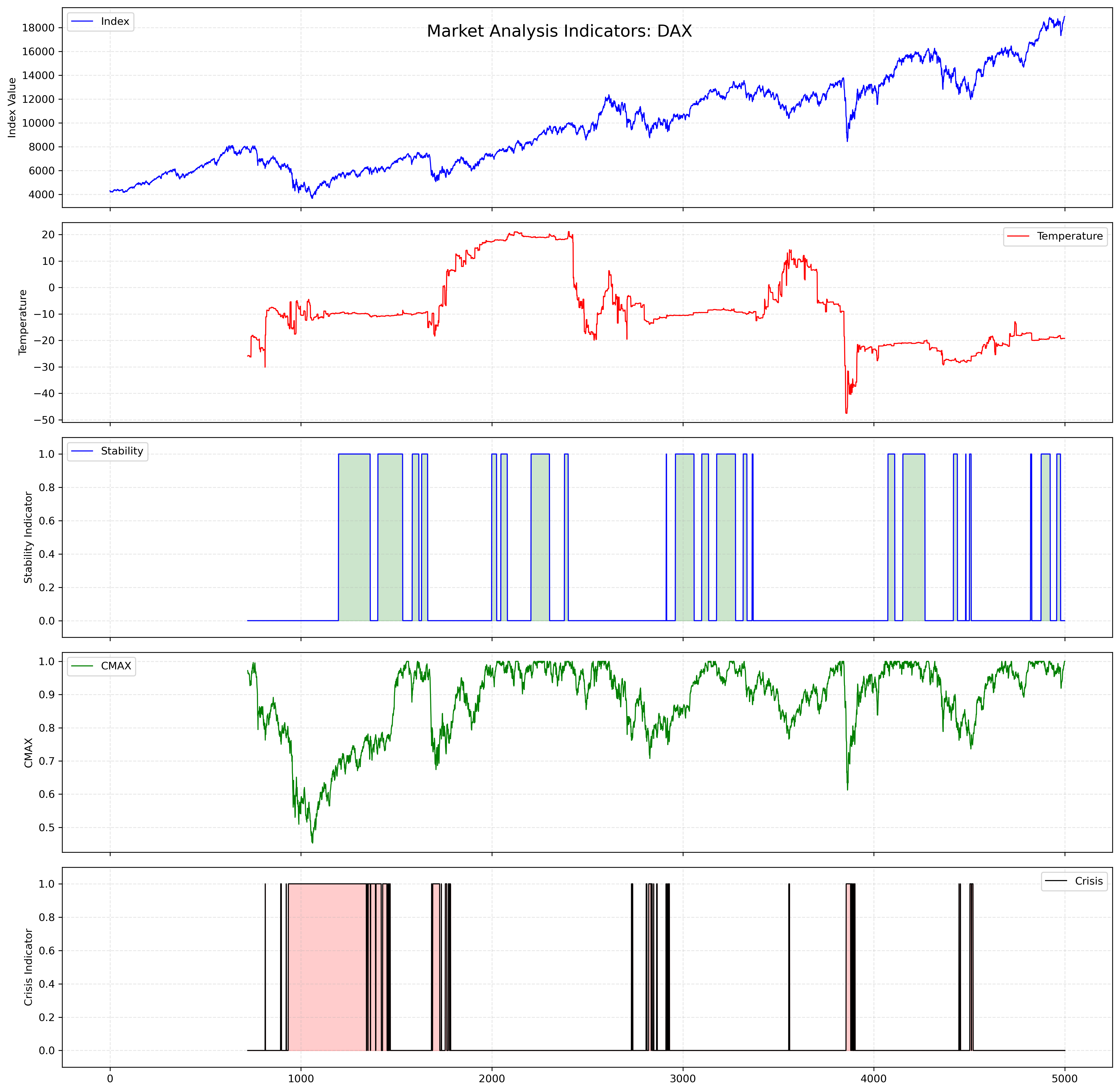}
	\caption{Market Analysis Indicators for S\&P 500. From top to bottom: (1) Index value over time, (2) Market temperature ($\Delta\beta$), (3) Stability indicator showing periods of temperature stability (blue shading), (4) CMAX values showing market drawdowns, and (5) Crisis indicator showing identified crisis periods (red shading).}
	\label{fig:DAX_analysis}
	\clearpage
\end{figure}

\newpage

\subsection*{A5. FTSE 100}
\begin{figure}[h]
	\centering
	\includegraphics[width=\textwidth]{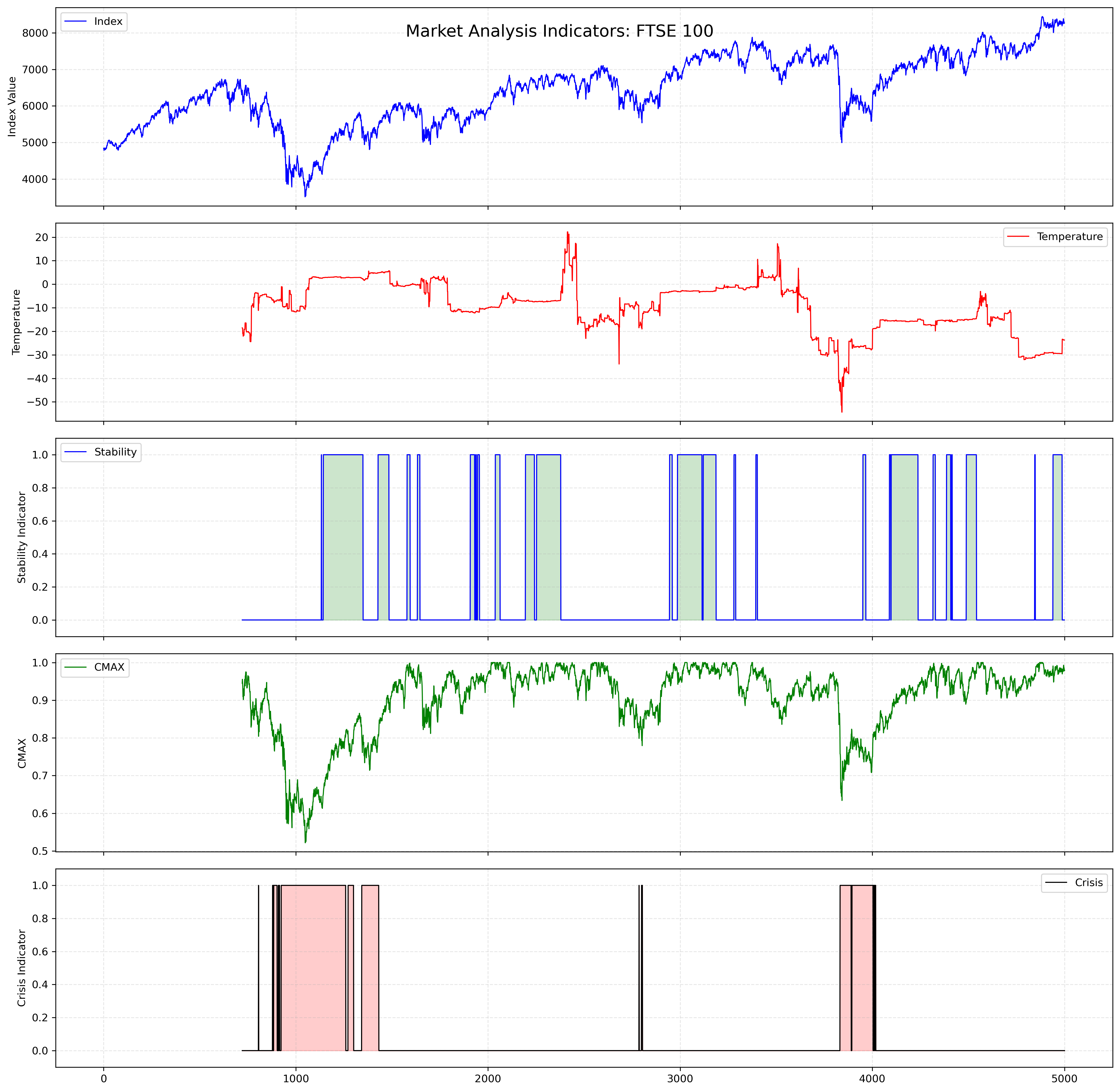}
	\caption{Market Analysis Indicators for S\&P 500. From top to bottom: (1) Index value over time, (2) Market temperature ($\Delta\beta$), (3) Stability indicator showing periods of temperature stability (blue shading), (4) CMAX values showing market drawdowns, and (5) Crisis indicator showing identified crisis periods (red shading).}
	\label{fig:FTSE100_analysis}
	\clearpage
\end{figure}

\newpage

\subsection*{A6. Euro Stoxx 50}
\begin{figure}[h]
	\centering
	\includegraphics[width=\textwidth]{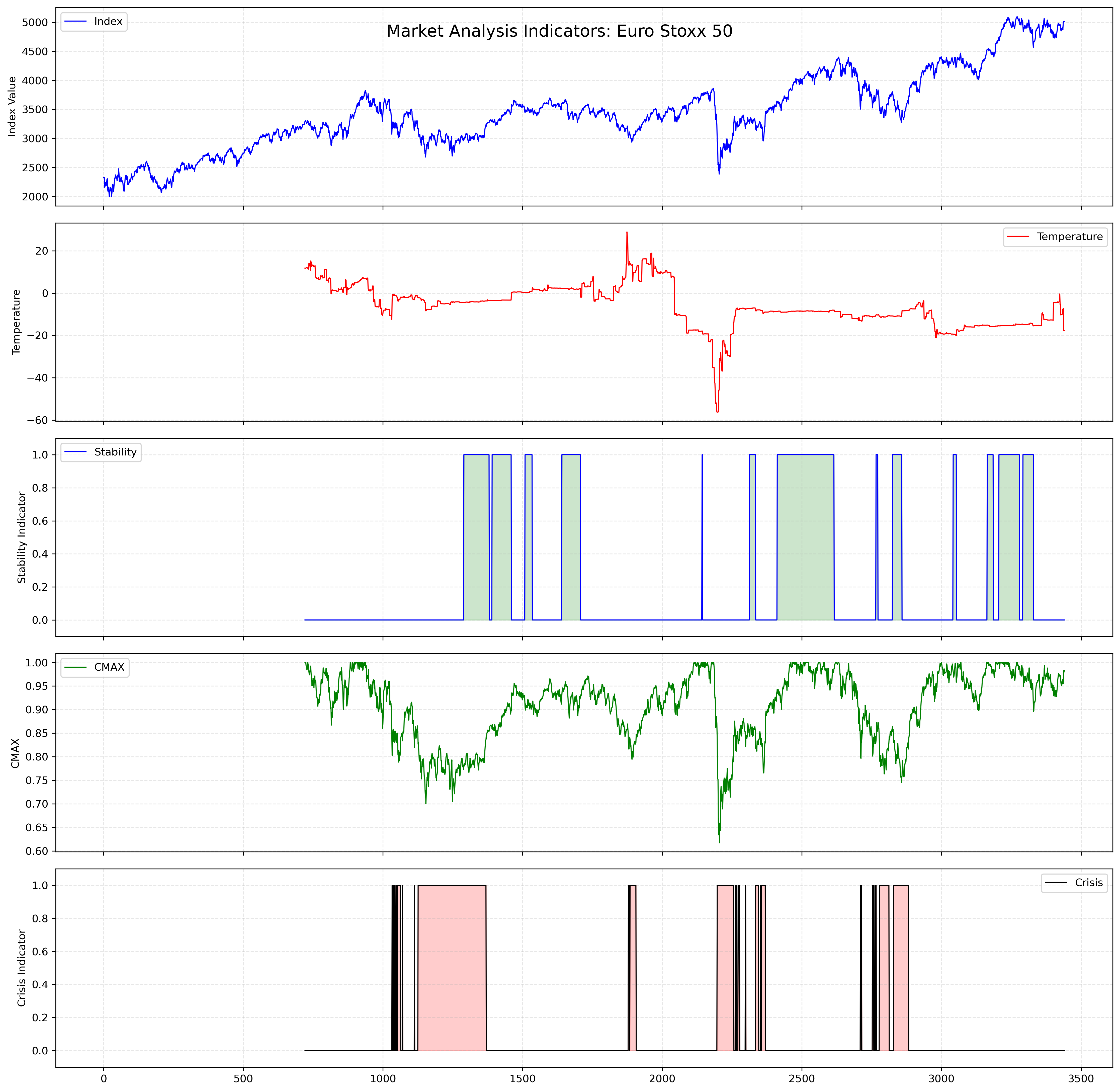}
	\caption{Market Analysis Indicators for S\&P 500. From top to bottom: (1) Index value over time, (2) Market temperature ($\Delta\beta$), (3) Stability indicator showing periods of temperature stability (blue shading), (4) CMAX values showing market drawdowns, and (5) Crisis indicator showing identified crisis periods (red shading).}
	\label{fig:Euro_Stoxx50_analysis}
	\clearpage
\end{figure}

\newpage

\subsection*{A7. Hang Seng}
\begin{figure}[h]
	\centering
	\includegraphics[width=\textwidth]{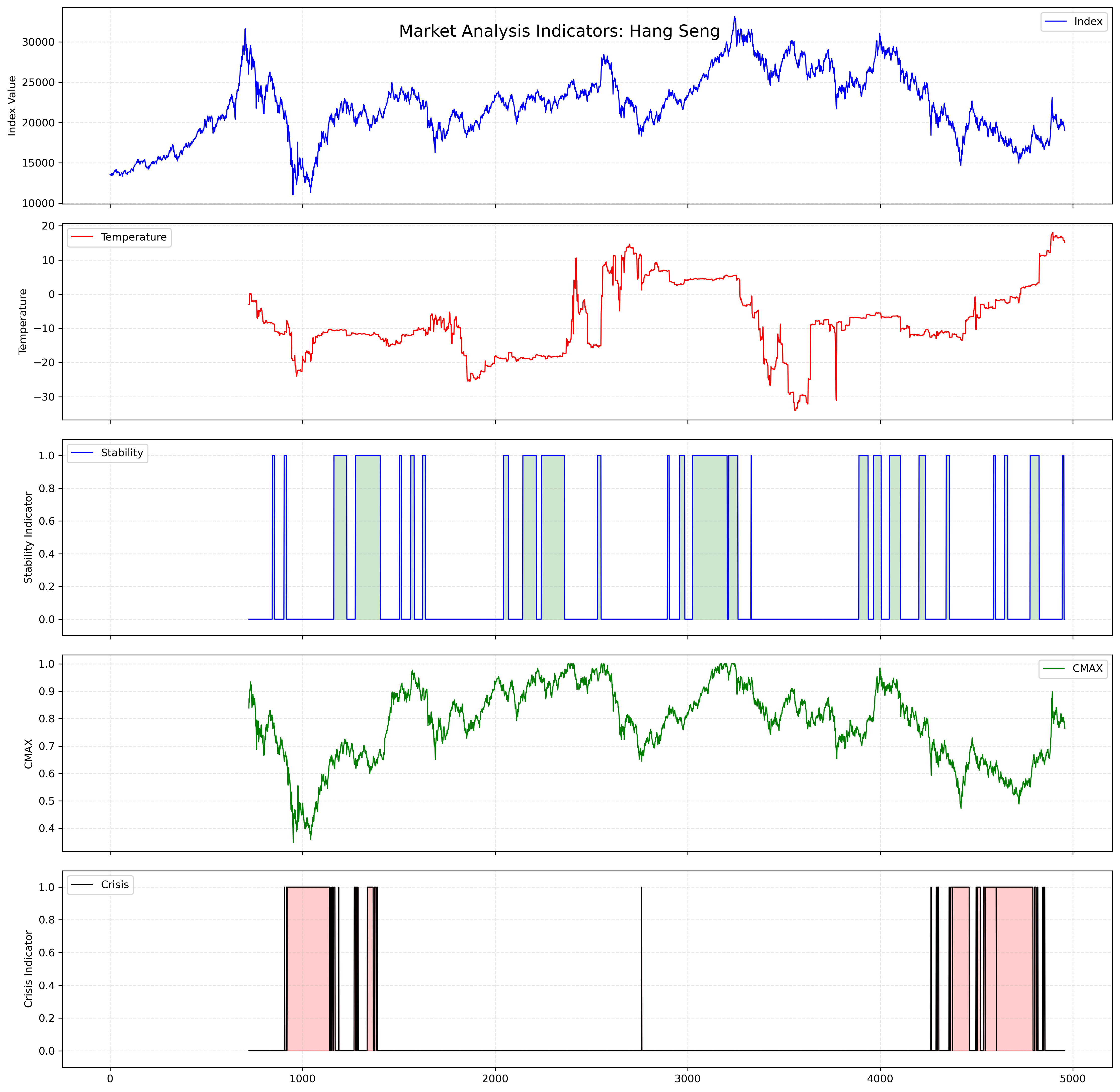}
	\caption{Market Analysis Indicators for S\&P 500. From top to bottom: (1) Index value over time, (2) Market temperature ($\Delta\beta$), (3) Stability indicator showing periods of temperature stability (blue shading), (4) CMAX values showing market drawdowns, and (5) Crisis indicator showing identified crisis periods (red shading).}
	\label{fig:HangSeng_analysis}
	\clearpage
\end{figure}

\newpage

\subsection*{A8. S\&P/ASX 200}
\begin{figure}[h]
	\centering
	\includegraphics[width=\textwidth]{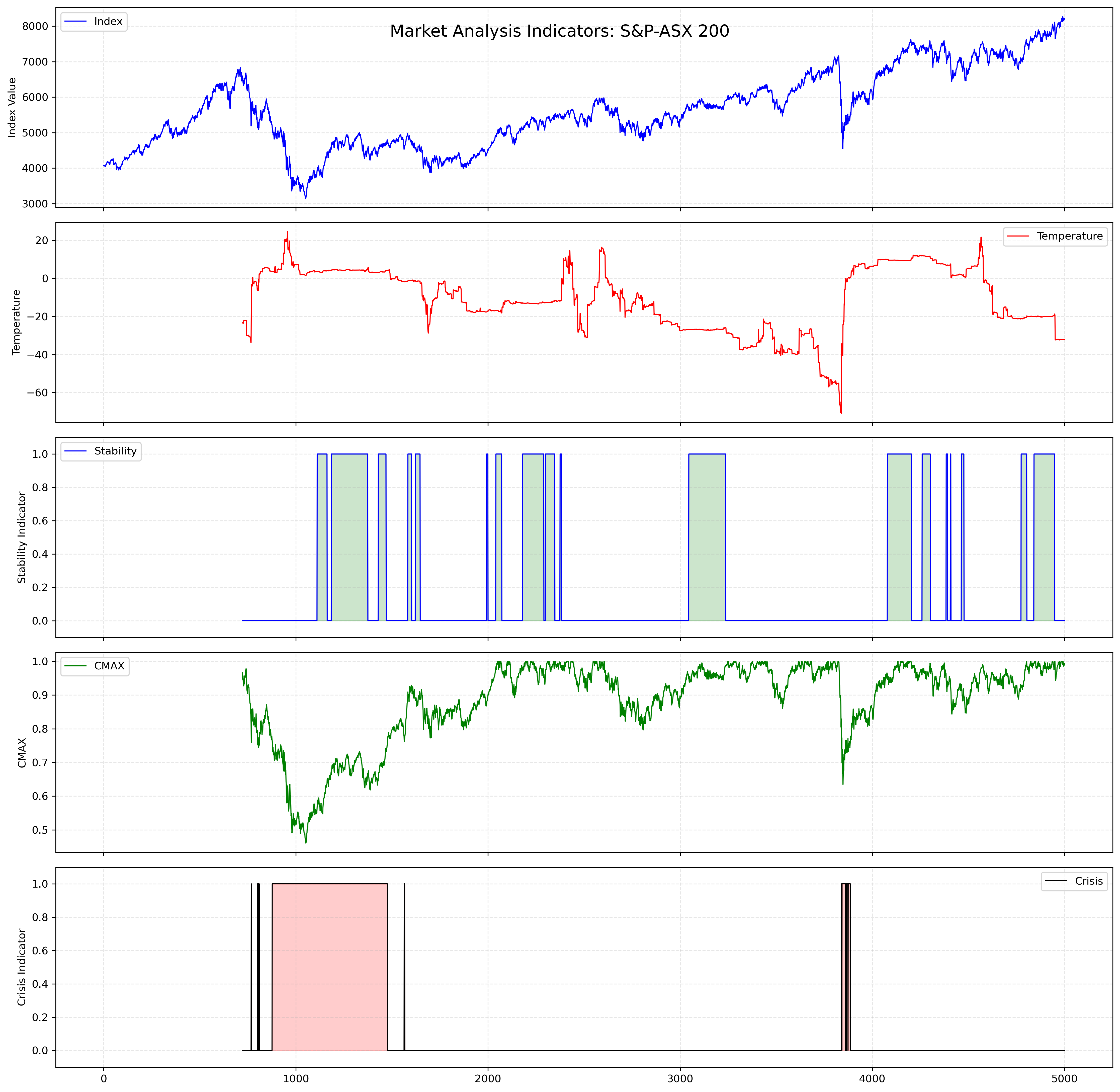}
	\caption{Market Analysis Indicators for S\&P 500. From top to bottom: (1) Index value over time, (2) Market temperature ($\Delta\beta$), (3) Stability indicator showing periods of temperature stability (blue shading), (4) CMAX values showing market drawdowns, and (5) Crisis indicator showing identified crisis periods (red shading).}
	\label{fig:S&P/ASX 200_analysis}
	\clearpage
\end{figure}

\newpage

\subsection*{A9. Nifty 50}
\begin{figure}[h]
	\centering
	\includegraphics[width=\textwidth]{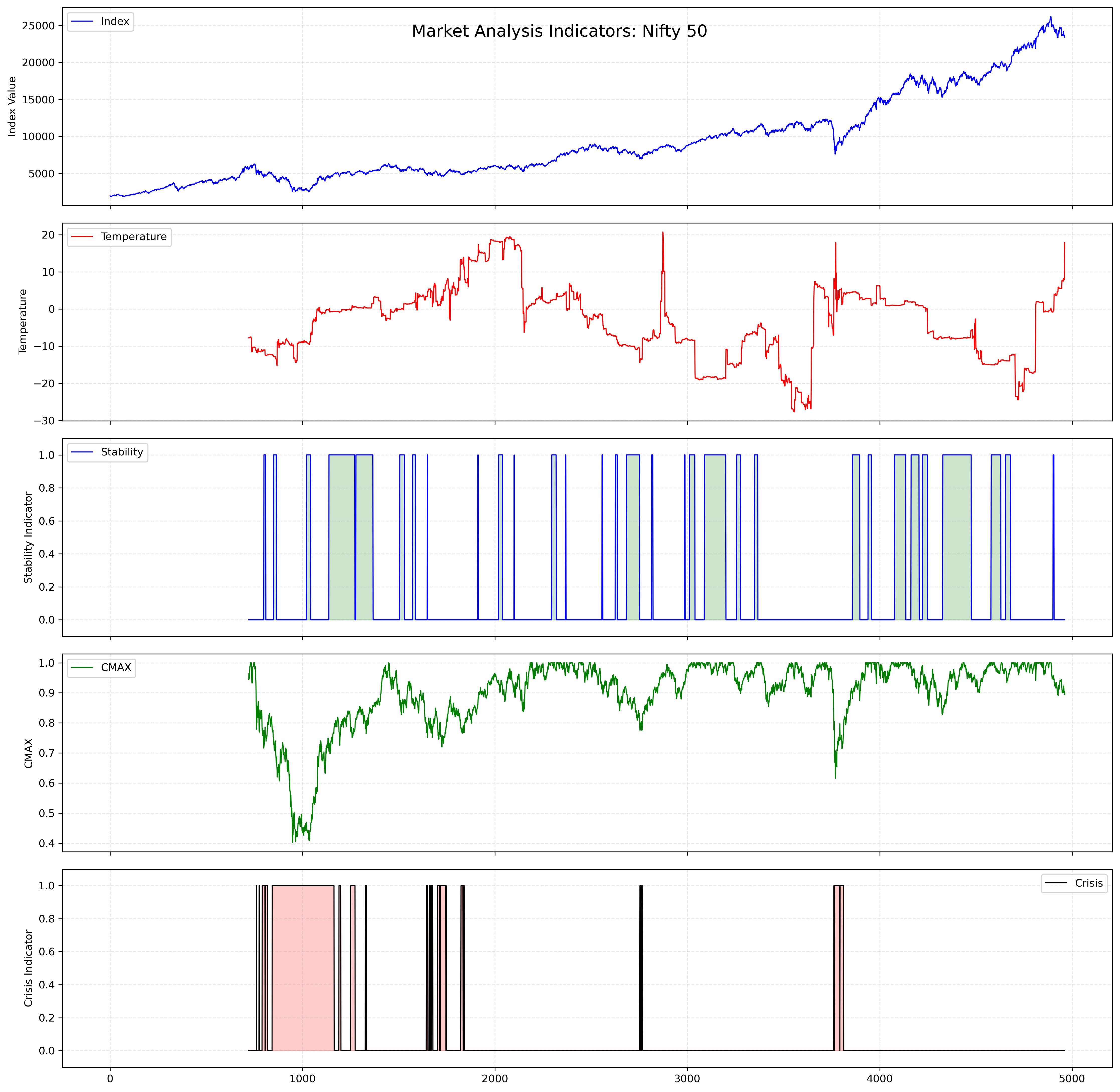}
	\caption{Market Analysis Indicators for S\&P 500. From top to bottom: (1) Index value over time, (2) Market temperature ($\Delta\beta$), (3) Stability indicator showing periods of temperature stability (blue shading), (4) CMAX values showing market drawdowns, and (5) Crisis indicator showing identified crisis periods (red shading).}
	\label{fig:Nifty50_analysis}
	\clearpage
\end{figure}



\end{document}